\documentclass{aa}  
\pdfoutput=1
\usepackage{graphicx}
\usepackage{natbib}
\usepackage{rotating}
\usepackage{txfonts}
\usepackage{xcolor}

\begin{document}

   \title{Bar pattern speeds in CALIFA galaxies}

   \subtitle{II. The case of weakly barred galaxies}

   \author{Virginia Cuomo
          \inst{1}
          \and J. Alfonso L. Aguerri\inst{2,3}
          \and Enrico Maria Corsini \inst{1,4}
          \and Victor P. Debattista\inst{5}
          \and Jairo M\'endez-Abreu
          \inst{2,3}
          \and Alessandro Pizzella
          \inst{1,4}
          }

   \institute{Dipartimento di Fisica e Astronomia ”G. Galilei”, Università di Padova, vicolo dell'Osservatorio 3, I-35122 Padova, Italy\\
              \email{virginia.cuomo@phd.unipd.it}
         \and
             Departamento de Astrof\'isica, Universidad de La Laguna, Avenida Astrof\'isico Francisco S\'anchez s/n, 38206 La Laguna, Tenerife, Spain
        \and
            Instituto de Astrof\'isica de Canarias, calle V\'ia L\'actea s/n, 38205 La Laguna, Tenerife, Spain
        \and
            INAF - Osservatorio Astronomico di Padova, vicolo dell'Osservatorio 2, I-35122 Padova, Italy
        \and
            Jeremiah Horrocks Institute, University of Central Lancashire, PR1 2HE Preston, UK
         }

   \date{Received 31 July, 2019; accepted 2 September, 2019}

 
  \abstract
   {About $35\%$ of the nearby disc galaxies host a weak bar for which different formation scenarios including the weakening of a strong bar, and tidal interaction with a companion, have been suggested. Measuring the bar pattern speeds of a sample of weakly-barred galaxies is a key step to constrain their formation process, but such a systematic investigation is still missing.}
   {We aim at investigating the formation process of weak bars by measuring their properties in a sample of 29 nearby weakly barred galaxies, spanning a wide range of morphological types and luminosities. The sample galaxies were selected to have an intermediate inclination, a bar at an intermediate angle between the disc minor and major axes, and an undisturbed morphology and kinematics to allow the direct measurement of the bar pattern speed. Combining our analysis with previous studies, we compared the properties of weak and strong bars.} 
   {We measured the bar radius and strength from the $r$-band images available in the Sloan Digital Sky Survey and bar pattern speed and corotation radius from the stellar kinematics obtained by the Calar Alto Legacy Integral Field Area Survey. We derived the bar rotation rate as the ratio between the corotation and bar radii.}
   {Thirteen out of 29 galaxies ($45\%$), which were morphologically classified as weakly barred from a visual inspection, do not actually host a bar component or their central elongated component is not in rigid rotation. We successfully derived the bar pattern speed in 16 objects. Two of them host an ultrafast bar. Using the bar strength to differentiate weak and strong bars, we found that the weakly-barred galaxies host shorter bars with smaller corotation radii than their strongly barred counterparts. Weak and strong bars have similar bar pattern speeds and rotation rates, which are all consistent with being fast. We did not observe any difference between the bulge prominence in weakly and strongly-barred galaxies, whereas nearly all the weak bars reside in the disc inner parts, contrary to strong bars.}
   {We ruled out that the bar weakening is only related to the bulge prominence and that the formation of weak bars is triggered by the tidal interaction with a companion. Our observational results suggest that weak bars may be evolved systems exchanging less angular momentum with other galactic components than strong bars.}


   \maketitle
%

\section{Introduction}

Bars are so common in the centre of disc galaxies that in the most-widely adopted morphological classifications \citep{Hubble1936, deVaucouleurs1959, vandenBergh1976}, barred galaxies represent one of the main families of both lenticular and spiral galaxies. Since bars show a wide variety of properties in terms of size, luminosity, and shape, disc galaxies are divided into unbarred, weakly, and strongly barred galaxies. 

Now it is known that bars are hosted in $\sim50\%$ of galactic discs in the local universe if observed in the optical bands, and this fraction rises to $\sim70\%$ in the near-infrared \citep{Aguerri2009, Buta2015} and about half of them are classified as weak bars \citep{Buta2007}. The bar fraction in nearby galaxies depends on the morphological type \citep{Marinova2007, Aguerri2009, Li2017} and is a strong function of the galaxy luminosity (or equivalently stellar mass), since it peaks for giant galaxies and decreases in both the low and high-mass regimes \citep{MendezAbreu2010, Nair2010, SanchezJanseen2010}. The bar fraction distribution as a function of galaxy luminosity varies significantly from cluster to field environments \citep{Barway2011, MendezAbreu2012, Lin2014}. A very weak trend is found between bar fraction and colour, slowly declining to redder colours \citep{Barazza2008}, although there are conflicting results \citep{Masters2011,Erwin2018}.

The radius, $R_{\rm bar}$, strength, $S_{\rm bar}$, and pattern speed, $\Omega_{\rm bar}$ are the main properties of a bar. The bar radius defines the extension of the stellar orbits supporting the bar, which mainly belong to the highly-elongated $x_1$ family \citep{Contopoulos1980, Manos2011}, even if a fraction of stochastic orbits is also theoretically predicted \citep{Martinet1990, Patsis2014}. The bar strength measures the non-axisymmetric forces produced by the bar potential \citep{Laurikainen2002}. The bar pattern speed is the angular frequency with which the bar rotates around the galactic centre \citep{Athanassoula2003, Combes2011}. This parameter is related with both the light and mass distribution of the host galaxy since it mainly depends on the redistribution of angular momentum between the galactic components \citep{Debattista2000, Athanassoula2003}. The bar pattern speed is usually parametrised by the bar rotation rate ${\cal R}=R_{\rm cr}/R_{\rm bar}$, where $R_{\rm cr}$ is the corotation radius, where the gravitational attraction balances the centrifugal acceleration in the rest-frame of the bar. Bars having $1.0\leq {\cal R}\leq 1.4$ end close to corotation and they are called {\em fast}. In contrast, bars with ${\cal R} >1.4$ are shorter than corotation and they are termed {\em slow} \citep{Debattista1998, Debattista2000, Athanassoula2002}. Bars with ${\cal R}<1.0$ corresponds to an unphysical regime for the $x_1$ orbits \citep{Contopoulos1981, Athanassoula1992,Vasiliev2015} and are called {\em ultrafast} \citep[][hereafter Paper I]{Buta2009,Aguerri2015}. 

The formation of a bar in an isolated galaxy is generally attributed to internal processes and typically includes three main phases: the initial growth, subsequent buckling, and final secular evolution \citep[e.g.][]{Hohl1971,Noguchi1987,Sellwood1981,Toomre1981,Raha1991,Debattista2006, Athanassoula2013,MartinezValpuesta2017}. The bar growth takes $\sim2$ Gyr, at the end of which a clear non-axysimmetric stellar structure stands out in the disc. During the buckling phase, which lasts $\sim1$ Gyr, the bar weakens. The following secular evolution takes place during several Gyrs and the bar slowly increases its length and strength. The bar pattern speed decreases at a rate depending on the amount of angular momentum exchanged between the disc and other galactic components  and on the dynamical friction exerted on the bar by the dark matter (DM) halo. Measuring the bar rotation rate allows to investigate the bar evolution and at the same time constrain the DM distribution \citep{Debattista2000, Athanassoula2003}. Other external events, such as interactions with companions and satellites \citep{Athanassoula2013, MartinezValpuesta2016, Lokas2018}, and internal processes such the gas fraction, the shape of DM halo, and the presence of a central mass concentration \citep{Athanassoula2003, Debattista2006, Athanassoula2013} further influence the formation and evolution of a bar. More specifically, \cite{Bournaud2002} and \cite{Bournaud2005} found that a significant gas accretion in the presence of a massive bulge produces different episodes of bar destruction and rebuilding. At each step, the newly formed bar is shorter and weaker while its pattern speed is faster than the previous one. 

A first pioneering effort to explain the formation of weak bars was done by \cite{Kormendy1979}, who concluded that lenses are the end result of the evolution of bars into nearly-axisymmetric structures. Since the fraction of barred galaxies hosting a lens is very high, the mechanism should be secular and possibly involve the interaction with the bulge. Moreover the majority of lenses are located in early-type galaxies with large central concentrations. This is in conflict with the findings of \cite{Laurikainen2013}, who found that the radius of fully developed lenses is on average $\sim1.3$ times larger than that of bars. Recently, \cite{Kruk2018} have provided some evidence supporting the idea that bars dissolve into lenses. They found that unbarred discs are bluer than their barred counterparts while unbarred galaxies with a lens are similar to strongly barred galaxies. Weakly barred galaxies are very similar to unbarred galaxies since their discs are bluer and their bars are shorter than those in strongly barred galaxies \citep{Abraham1999}. Weak bars can also represent the end result of weak interactions. \cite{MartinezValpuesta2017} investigated, through numerical simulations, the formation of bars triggered or affected by fast interactions. These bars formed by interactions are slow throughout their lifetime. Low values of the bar pattern speed (corresponding to ${\cal R}\sim2$) have been also found by \cite{Lokas2018} in the late evolutionary stages of tidally induced bars. 

Therefore measuring the bar pattern speed of a sample of weakly barred galaxies could constrain their formation process. This is a challenging task; in the past a variety of indirect methods based on modelling has been used to measure the pattern speed and corresponding rotation rates of strong bars. The only model-independent method proposed so far is the one by \citet[][hereafter TW]{Tremaine1984}. In the last two decades, the applications of the TW method using long-slit spectroscopic data of stellar kinematics allowed to study $\sim20$ galaxies \citep[see ][for a review]{Corsini2011}. More recently, integral-field spectroscopy have been shown to remarkably improve the efficiency and precision of the TW measurements. In Paper I, we measured the bar pattern speed of 15 galaxies on the stellar velocity maps provided by Calar Alto Legacy Integral Field Area survey \citep[CALIFA,][]{sanchez2012} and \cite{Guo2019} obtained the bar pattern speed for another 51 galaxies using the integral-field spectroscopic data from the Massive Nearby Galaxies survey \citep[MaNGA,][]{Bundy2015}. However, neither Paper I nor \citet{Guo2019} included weakly barred galaxies in their samples. To date, the bar pattern speed was measured with the TW method only in one weakly barred galaxy, ESO~139-G0009, which turned out to host a fast bar \citep{Aguerri2003}. The bar pattern speed has been indirectly measured from the velocity field of the ionized/molecular gas in a number of weakly barred galaxies suggesting that slow bars are hosted especially by late-type spirals, in spite of large uncertainties on $\cal{R}$ \citep{Hirota2009, Font2017, Salak2019}.

\begin{figure*}
    \centering
    \includegraphics[scale=0.7]{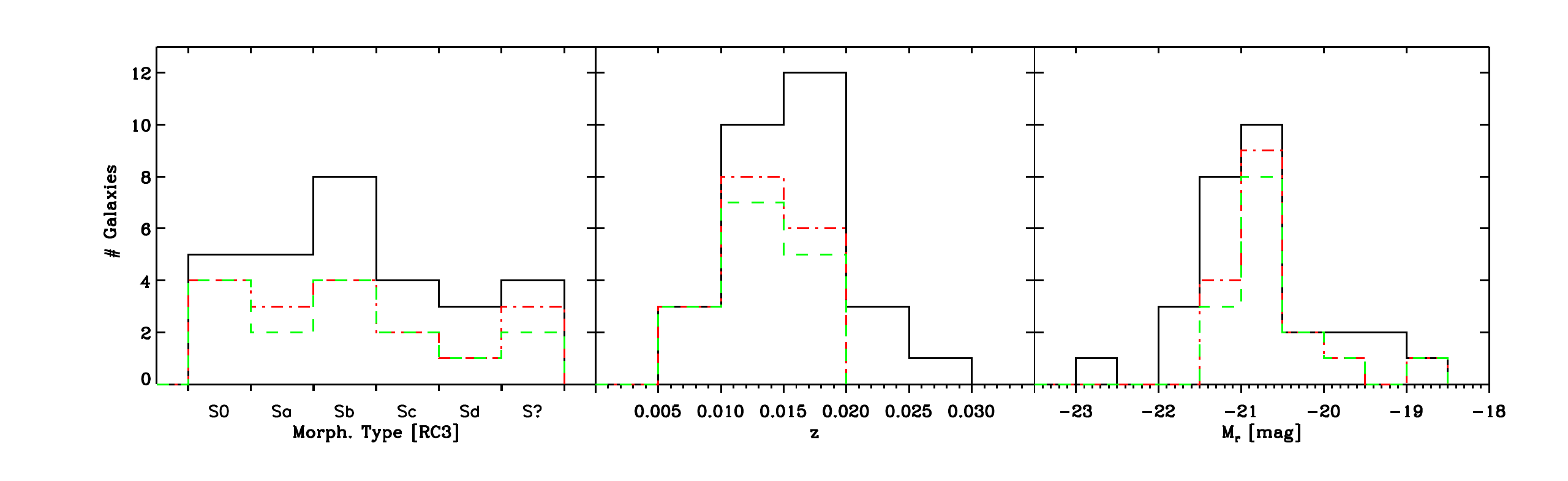}
    \caption{Distribution of the morphological type ({\em left panel}), redshift ({\em central panel}), and absolute $r$-band magnitude ({\em right panel}) of our sample of 29 {\em bona fide} SAB galaxies (black solid line), 
    the sample of 16 SAB galaxies successfully analysed with the TW method (red dot-dashed line), and of the 14 SAB galaxies hosting a non-ultrafast bar (green dashed line).}
    \label{fig:histograms}
\end{figure*}

In this paper we aim at investigating the formation of weak bars by measuring the bar radius, strength, and pattern speed with the TW method in a sample of weakly barred galaxies for which integral-field spectroscopic data are available from the CALIFA survey. The paper is structured as follows. We present the galaxy sample in Sec.~\ref{sec:sample}. We measure the bar properties of the sample galaxies in Sec.~\ref{sec:bar}. We present our results in Sec.~\ref{sec:results}. We discuss our conclusions in Sec.~\ref{sec:conclusions}. We adopt as cosmological parameters $\Omega_{\rm m} = 0.286$, $\Omega_{\Lambda} = 0.714$, and $H_0=69.3$ km s$^{-1}$ Mpc$^{-1}$ \citep{Hinshaw2013}.

\begin{sidewaystable*}
\caption{\label{tab:sample} Properties of the {\em bona fide} SAB galaxies.}
\begin{tabular}{ccccccccccccccc}
\hline\hline
Galaxy & Alt. Name & Morph. Type & Morph. Type & $z$ & $m_r$ & $D$ & $M_r$ & $D_{25}\times d_{25}$ & $\Delta{\rm PA}$ & $i$ & Rejection & $R_{\rm bar}$ & $S_{\rm bar}$\\
 & & [RC3] & [CALIFA] & & [mag] & [Mpc] & [mag] & [arcmin] & [$^\circ$] & [$^\circ$] & & [kpc] & \\
 (1) & (2) & (3) & (4) & (5) & (6) & (7) & (8) & (9) & (10) & (11) & (12) & (13) & (14)\\ 
\hline
IC 1199 & PGC 57373 & S? & SABb & 0.016 & 14.33 & 69.5 & $-19.88$ & $1.35\times0.46$ & 11.6 & $67.0\pm0.5$  & TW & $7.16^{+3.79}_{-2.51}$ &  $0.37^{+0.03}_{-0.04}$ \\
IC 1256& PGC 60203 & Sb & SABb & 0.016 & 13.60 & 67.7 & $-20.55$ & $1.58\times1.15$ & 50.0 & $51.3\pm0.7$ & TW & 2.71$^{+0.22}_{-0.24}$ & 0.21$^{+0.01}_{-0.02}$\\
IC 1528& PGC 312& SB(r)b? & SABbc& 0.013 & 12.92 & 49.9 & $-20.57$ & $2.40\times1.15$ & 57.5 & $66.7\pm0.8$ &... & 2.15$^{+0.66}_{-0.71}$ & 0.235$^{+0.002}_{-0.016}$\\
IC 1683& PGC 5008 & S? & SABb & 0.016 & 13.37 & 66.2 & $-20.73$ & $1.32\times0.60$ & 35.5 & $54.3\pm0.9$ & ...& 8.79$^{+0.62}_{-0.65}$ & 0.73$^{+0.07}_{-0.08}$\\
IC 5309& PGC 71051 & Sb & SABc & 0.014 & 13.72 & 55.2 & $-19.99$ & $1.35\times0.60$ & 10.5 & $60.0\pm1.4$ &... & $1.98^{+0.89}_{-0.50}$ & $0.205^{+0.006}_{-0.019}$\\
MCG-02-02-030& PGC 1841& (R)SB(r)a & SABb & 0.012 & 12.74 & 45.9 & $-20.57$ & $1.95\times0.83$ & 37.5 & $68.7\pm0.5$ & ...& 3.6$^{+2.3}_{-1.2}$ & 0.28$^{+0.03}_{-0.03}$\\
NGC 192& PGC 2352& (R')SB(r)a: & SABab & 0.014 & 12.39 & 54.7 & $-21.30$ & $1.91\times0.89$ & 73.7 & $70.5\pm0.5$ & ...& 11. 0$^{+1.8}_{-1 .4}$ & 0.83$^{+0.09}_{-0.09}$\\
NGC 234& PGC 2600& SAB(rs)c & SABc & 0.015 & 12.50 & 59.5 & $-21.37$ & $1.62\times1.62$ & 20.0 & $29.4\pm0.4$ & Fourier & ... & ...\\
NGC 364& PGC 3833 & (R)SB(s)0: & EAB7 & 0.017 & 12.92 & 69.3 & $-21.28$ & $1.41\times1.26$ & 57.8 & $44\pm1$ & ...& 3.17$^{+0.62}_{-0.64}$ & 0.46$^{+0.01}_{-0.02}$\\
NGC 477 & PGC 4915 & SAB(s)c & SABbc & 0.020 & 13.44 & 81.0 & $-21.10$ & $2.19\times1.17$ & 35.3 & $52\pm1$ & TW & $7.40^{+2.37}_{-1.34}$ & $0.427_{-0.004}^{+0.008}$\\
NGC 551 & PGC 5450 & SBbc & SABbc & 0.017 & 13.28 & 71.1 & $-20.98$ & $1.82\times0.79$ & 28.0 & $64.7\pm0.8$ & ... & $3.9^{+2.0}_{-2.1}$ & $0.30_{-0.07}^{+0.07}$\\
NGC 2410 & PGC 21336 & SBb? & SABb & 0.016 & 12.57 & 69.8 & $-21.65$ & $2.45\times0.71$ & 14.4 & $73.8\pm0.7$ & TW & $5.77^{+3.37}_{-2.16}$ & $0.55_{-0.07}^{+0.05}$\\
NGC 2449 & PGC 21802 & Sab & SABab & 0.016 & 12.87 & 73.3 & $-21.40$ & $1.35\times0.65$ & 59.2 & $69\pm2$ & ...& 4.59$^{+0.75}_{-0.80}$ & 0.60$^{+0.04}_{-0.03}$\\
NGC 2553& PGC 23240& S? & SABab & 0.016 & 13.04 & 71.5 & $-21.23$ & $0.87\times0.68$ & 48.1 & $54.6\pm0.9$ & ...& 7.7$^{+2.1}_{-1.8}$ & 0.57$^{+0.01}_{-0.01}$ \\
NGC 2880& PGC 26939 & SB0$^{-}$ & EAB7 & 0.0051 & 11.57 & 24.1 & $-20.34$ & $2.04\times1.20$ & 61.7 & $56.7\pm0.2$ &... & 1.49$^{+0.71}_{-0.42}$ & 0.452$^{+0.004}_{-0.006}$\\
NGC 3994 & PGC 37616 & S(r)c pec & SABbc & 0.010 & 12.68 & 48.6 & $-20.75$ & $1.05\times0.59$ & 80.0 & $63.0\pm0.5$ &... & 1.77$^{+0.56}_{-0.47}$ & 0.38$^{+0.01}_{-0.09}$\\
NGC 5056& PGC 46180 & Scd: & SABc & 0.019 & 13.13 & 84.3 & $-21.50$ & $1.74\times1.00$ & 34.6 & $59.1\pm0.8$ & TW & 2.25$^{+0.18}_{-0.31}$ & 0.40$^{+0.05}_{-0.03}$\\
NGC 5971& PGC 55529& Sa & SABb &0.011 & 13.42 & 62.8 & $-20.57$ & $1.58\times0.56$ & 23.8 & $69.0\pm0.5$ & ...& 7.3$^{+6.1}_{-3.3}$ & 0.504$^{+0.009}_{-0.004}$\\
NGC 6278& PGC 59426 & S0 & SAB0/a & 0.0090 & 12.20 & 40.9 & $-20.86$ & $2.04\times1.20$ & 10.7 & $58.8\pm0.3$ &... & 2.84$^{+1.09}_{-0.17}$ & 0.36$^{+0.04}_{-0.04}$\\
NGC 6427& PGC 60758 & S0$^{-}$: & SAB0 & 0.011 & 12.56 & 45.7 & $-20.74$ & $1.58\times0.63$ & 44.1 & $68.1\pm0.7$ &... & 1.9$^{+1.7}_{-1.0}$ & 0.63$^{+0.02}_{-0.02}$\\
NGC 6978 & PGC 65631 & Sb & SABb & 0.020 & 12.84 & 82.8 & $-21.75$ & $1.48\times0.69$ & 10.0 & $68.5\pm0.8$ & Fourier & ... & ... \\
NGC 7787 & PGC 72930 & (R')SB(rs)0/a: & SABab & 0.022 & 13.91 & 93.0 & $-20.88$ & $1.78\times0.47$ & 46.7 & $65\pm1$  & Fourier & ... & ... \\
UGC 36 & PGC 366 & S(r)a: & SABab & 0.021 & 13.27 & 85.6 & $-21.39$ & $1.32\times0.60$ & 63.9 & $63.0\pm0.8$ & TW & $6.80_{+1.41}^{-1.41}$ & $0.43_{-0.04}^{+0.04}$\\
UGC 3944& PGC 21475 & Scd: & SABbc & 0.0083 & 13.80 & 58.3 & $-20.03$ & $1.78\times0.79$ & 79.8 & $59.3\pm0.5$ & ...& 1.87$^{+0.88}_{-0.63}$ & 0.27$^{+0.05}_{-0.02}$\\
UGC 7012& PGC 37976 & Scd: & SABcd & 0.010 & 14.06 & 49.7 & $-19.37$ & $1.66\times0.89$ & 10.0 & $56.5\pm0.8$ & Fourier & ... & ...\\
UGC 8231& PGC 45561 & SB? & SABd & 0.0083 & 14.18 & 37.8 & $-18.71$ & $1.58\times0.56$ & 13.0 & $68.1\pm0.5$ &... & 2.30$^{+0.47}_{-0.70}$ & 0.24$^{+0.04}_{-0.04}$\\
UGC 9067& PGC 50621 & Sab & SABbc & 0.026 & 13.66 & 116.6 & $-21.67$ & $1.38\times0.68$ & 14.0 & $63\pm1$ & TW & $7.80^{+3.62}_{-2.30}$ & $0.25^{+0.04}_{-0.06}$\\
UGC 10693 & PGC 59560 & E & EAB7 & 0.015 & 12.51 &122.9 & $-22.89$ & $1.82\times1.23$ & 10.0 & $49.8\pm0.6$ & Fourier & ... & ...\\	
UGC 10796 & PGC 59997 & SB(s)b & SABcd & 0.010 & 13.91 & 43.8 & $-19.30$ & $1.58\times1.20$ & 66.5 & $57\pm1$ & Fourier & ... &...\\
\hline
\end{tabular}
\tablefoot{(1) Galaxy name. (2) Alternative name. (3) Morphological type from RC3. (4): Morphological type from \cite{Walcher2014}. (5) Redshift from SDSS-DR14 \citep{Abolfathi2018}. (6) Apparent model total $r$-band magnitude from SDSS-DR14. (7) Distance from NASA/IPAC Extragalactic Database (NED)\footnote{Available at \url{https://ned.ipac.caltech.edu/}}. (8) Absolute SDSS $r$-band magnitudes $M_r$ obtained as described in Sec.~\ref{sec:sample}. (9) Major and minor diameters of the isophote at a surface brightness level  of $\mu_B=25$ mag arcsec$^{-2}$ from \cite{deVaucouleurs1991}. (10) Difference between the PA of the bar and disc major axis either obtained from the isophototal analysis of the $r$-band SDSS image or provided by \cite{MendezAbreu2017}. (11) Galaxy inclination obtained from the disc ellipticity provided by \cite{MendezAbreu2017} and assuming an infinitesimally-thin disc. (12) Reason for the rejection of the galaxy, where Fourier signifies no evidence of a bar from the Fourier analysis of the $r$-band SDSS image, TW signifies unsuccessful application of the TW method to the CALIFA stellar velocity field. (13) Bar radius obtained as described in Sec.~\ref{sec:radius}. (14) Bar strength obtained as described in Sec.~\ref{sec:strength}. }
\end{sidewaystable*}

\section{Sample selection}
\label{sec:sample}

The aim of this work is to analyse the bar properties for a large sample of galaxies with a TW-measured $\Omega_{\rm bar}$ and spanning a wide range of bar strengths. Several strongly barred galaxies have been already measured in literature, but there is a lack of weak bars. To fill this gap we take galaxies from the CALIFA survey which aims at measuring the properties of a statistically significant sample of nearby galaxies with integral field spectroscopy \citep{sanchez2012}. The CALIFA Data Release 3 \citep[DR3,][]{Sanchez2016} includes $\sim 700$ galaxies from the SDSS Data Release 7 \citep[SDSS-DR7,][]{Abazajian2009}, which are selected to have a major-axis diameter $45 < D_{25} < 80$ arcsec in the $r$-band and a redshift $0.005 < z < 0.03$. They were observed with the Potsdam Multi Aperture Spectrograph \citep[PMAS,][]{Roth2005} mounted at the 3.5-m telescope of the Calar Alto Observatory \citep{Husemann2013, Walcher2014}. We considered the 265 CALIFA galaxies, which were morphologically classified as doubtful barred galaxies \citep{Walcher2014}. The visual identification of bars is not always obvious and it is even more difficult in the case of weakly barred galaxies. The morphological classification performed by the CALIFA collaboration does not always match the \citet[][hereafter RC3]{deVaucouleurs1991} classification (Table~\ref{tab:sample}). 

From the CALIFA SAB galaxies we selected those for which the stellar kinematic maps were measured by \cite{FalconBarroso2017}. These selection criteria allowed us to discard {\em a priori} the objects with a disturbed kinematics. We remained with 58 galaxies visually classified as SAB, for which we analysed the stellar kinematics obtained with a spectral resolution of $R=1650$ (corresponding to $\sigma_{\rm inst} \sim 70$ km~s$^{-1}$ at 4500 \AA) and a spatial resolution of 1 arcsec.

The TW method allows to measure $\Omega_ {\rm bar}$ from
\begin{equation}
\langle X\rangle \: \Omega_{\rm bar} \sin i = \langle V\rangle
\label{eq:tw}
\end{equation}
\noindent where $i$ is the disc inclination, and
\begin{equation}
\langle X\rangle=\frac{\int X \Sigma d\Sigma}{\int \Sigma d\Sigma} ;\;\; \langle V\rangle=\frac{\int V_{\rm los}\Sigma d\Sigma}{\int \Sigma d\Sigma}
\label{eq:integrals}
\end{equation}
\noindent are the photometric and kinematic integrals, defined as the luminosity-weighted average of position $X$ and line-of-sight (LOS) velocity $V_{\rm los}$, respectively, while $\Sigma$ represents the surface brightness of the galaxy. They have to be measured along directions parallel to the disc major axis. Then, fitting the values of $\langle X\rangle$ and $\langle V\rangle$ obtained for different offset positions crossing the bar with a straight line then gives $\Omega_ {\rm bar}\sin{i}$, so at least two pseudoslits has to be defined. The use of integral-field spectroscopic data allows measurement of $\langle X\rangle$ and $\langle V\rangle$ in several parallel pseudoslits by collapsing their corresponding spectra along the spectral and spatial directions, respectively. To this aim we followed the prescriptions of Paper I.

In order to apply any further analysis on our sample, we need to have the structural parameters of the galaxies. To this aim, we selected the 37 sample galaxies, whose SDSS $r$-band images were analysed by \cite{MendezAbreu2017}. They performed the isophotal analysis and photometric decomposition of the surface brightness distribution using the {\sc iraf} task {\sc ellipse} \citep{Jedrzejewski1987} and {\sc gasp2d} \citep{MendezAbreu2008, MendezAbreu2014}, respectively.

Moreover, to successfully apply the TW method, the galaxies should have an intermediate inclination and their bars should be elongated at an intermediate position angle (PA) between the disc major and minor axes. Low-inclination galaxies are characterised by small stellar velocities, large velocity errors, and a large uncertainty on the disc PA, while in highly-inclined galaxies it is difficult to identify the bar and to locate the pseudoslits. A bar aligned with the disc major axis gives $\langle X\rangle=0$ arcsec, while a bar aligned with the disc minor axis is characterised by $\langle V\rangle = V_{\rm sys}$. To address these issues we rejected all the galaxies with a $\Delta{\rm PA}<10^\circ$ between the bar major axis and disc major/minor axis and kept objects with a disc inclination $25^\circ<i<75^\circ$, as done in Paper I. When the photometric decomposition did not include a bar component, we recovered the bar PA from the analysis of the ellipticity $\epsilon$ and PA radial profiles as discussed in Sec.~\ref{sec:pa_ellipticity}. At the end, our sample of {\em bona fide} SAB galaxies totals 29 objects, whose main properties are listed in Table~\ref{tab:sample}. The distributions of their morphological types, redshifts, and absolute SDSS $r$-band magnitudes are plotted in Fig.~\ref{fig:histograms}. Since the distribution in morphologies, redshift and absolute magnitudes reflect the properties of the mother sample in CALIFA \citep{Walcher2014}, we can conclude there is no bias in our final selected sample with respect to the initial one.

In addition, the TW method works for a stellar tracer satisfying the continuity equation and because of this it was initially applied to early-type disc galaxies, which do not show strong evidence of spiral arms or heavily-patchy dust distribution. Spiral arms may lead to a wrong determination of the disc PA and their light contribution may affect the photometric integrals of the bar. The presence of dust and/or star formation may cause a non coincidence between the surface brightness and mass distribution of the galaxy, which results in a mismatch between photometric and kinematic measurements. These effects may be mitigated by computing the mass-weighted kinematic and photometric integrals \citep{Gerssen2007}. However, Paper I compared the values of $\Omega_{\rm bar}$ derived for a number of spirals from both light- and mass-weighted TW integrals and found consistent results, even in late-type galaxies, which are the most affected by this problem. This means that the TW method can be applied to barred spirals after checking the convergence of the TW integrals, which allows to control and limit contamination from other spurious features, such as foreground stars or bad pixels.  

\section{Properties of the weak bars}
\label{sec:bar}

\subsection{Disc inclination and position angle}
\label{sec:pa_ellipticity}

The TW method is very sensitive to the misalignment between the orientation of pseudoslits and $\langle V \rangle$ and disc major axis. In addition, the calculation of $\Omega_{\rm bar}$ also requires knowing the disc inclination \citep{Debattista2003}.

In order to accurately constrain the disc orientation, we decided to consider both the photometric decomposition and isophotal analysis of the SDSS $r$-band images provided by \cite{MendezAbreu2017}. For each object, we derived the disc inclination $i$ and PA from the ellipticity $\epsilon$ and major-axis PA of the ellipses fitting the outermost galaxy isophotes measured by \cite{MendezAbreu2017}. We defined the extension of the disc radial range by fitting the PA measurements with a straight line and selecting all the radii where the linear slope was consistent with zero within the associated root mean square error. We adopted the mean PA and mean $\epsilon$ and corresponding root mean square errors as the disc geometric parameters and their errors. Finally, we derived $i = \arccos{(1-\epsilon)}$ by assuming an infinitesimally thin disc. 

The resulting PA values are not always consistent within the $3\sigma$ errors with those from the photometric decomposition by \cite{MendezAbreu2017}. Although the mean difference of $\Delta {\rm PA}$ is lower than $1\fdg5$ in $80\%$ of the sample, for few galaxies the difference is as large as $\Delta {\rm PA}\sim7\degr$. It has to be noted that the bar component was not always included in the photometric decomposition due to the weakly barred nature of these galaxies and this affects the resulting best-fitting parameters of the disc \citep[see][for a discussion]{MendezAbreu2014,deLorenzoCaceres2019}. For each galaxy, we defined a range for the disc PA which covers the values from the photometric decomposition and isophotal analysis and their errors to be used for the application of the TW method. The $i$ values from the isophotal analysis are consistent with those from the photometric decomposition listed in Table~\ref{tab:sample}, which we adopt here.

\subsection{Bar detection}
\label{sec:fourier}

The visual identification of weakly barred galaxies is difficult \citep[e.g.][]{Nair2010, Lee2019} and most of our {\em bona fide\/} SAB galaxies were actually listed in RC3 either as unbarred galaxies or their classification was uncertain (Table~\ref{tab:sample}). For this reason, the accurate analysis of the galaxy surface brightness distribution is a mandatory step to identify the presence of a genuine bar component.

The Fourier analysis of the light distribution has been used often to detect and characterise the different galactic components, especially bars, which correspond to bisymmetric departures from axisymmetry \citep{Ohta1990, Athanassoula2003, GarciaGomez2017}. Following \cite{Aguerri2000}, we decomposed the deprojected azimuthal surface brightness profile $I(r,\phi)$ of each sample galaxy, where $(r,\phi)$ are the polar coordinates in the galaxy disc. We recovered the deprojected $r$-band image of the galaxy by stretching the original SDSS image along the disc minor axis by a factor equal to $1/\cos{i}$, with the flux conserved. To this aim, we adopted the geometric parameters of the disc derived by \cite{MendezAbreu2017}.

We know that the bar region is characterised by large values of the even Fourier components and in particular of the $m=2$ Fourier component. The odd Fourier components are generally smaller than the even ones because they are associated to the presence of asymmetric components. The maximum amplitude of the $m=2$ Fourier component is correlated with the bar strength and the bars typically have ${(I_{2}/I_{0})}_{\rm max}>0.2$ \citep{Aguerri2003}. We expect that the phase angle $\phi_2$ of the $m=2$ Fourier component is constant within the bar region. We found that the Fourier components of 6 sample galaxies do not meet these criteria. Therefore, we concluded that these galaxies are more likely to be unbarred systems and we excluded them from the analysis (Table~\ref{tab:sample}).

\subsection{Bar strength}
\label{sec:strength}

As in Paper I, we measured $S_{\rm bar} = {(I_{2}/I_{0})}_{\rm max}$ for all the weak bar sample galaxies as the maximum of the intensity ratio between the $m = 2$ and $m = 0$ Fourier components \citep{Athanassoula2002}. The uncertainties associated with the measurement of the strength are obtained by performing a Fourier analysis using the two portions of the deprojected azimuthal surface brightness $I(r, \phi)$ with $0\degr<\phi<180\degr$ and $180\degr<\phi<360\degr$. The difference between these two measurements with respect to the reference value obtained from the full surface brightness distribution provided the errors on the bar strength, typically smaller than $10\%$. The resulting values $S_{\rm bar}$ and corresponding errors are reported in Table~\ref{tab:sample}. We will define in Sec.~\ref{sec:weak_strong} a quantitative criterion based on $S_{\rm bar}$ to distinguish between weak and strong bars.

\subsection{Bar radius}
\label{sec:radius}

For each of the 23 galaxies confirmed to host a weak bar, we measured $R_{\rm bar}$ from the analysis of the SDSS $r$-band image using three independent methods as done in Paper I. We considered the bar-interbar intensity ratio obtained from the Fourier analysis, the location of the maximum in the $\epsilon$ radial profile, and the behaviour of the PA radial profile. The radial profiles of $\epsilon$ and PA are derived by fitting ellipses to the isophotes of the $r$-band image using {\sc ellipse} and considering a variable value for the centre of the galaxy \citep{MendezAbreu2017}.

First, we measured $R_{\rm bar}$ from the Fourier analysis by tracing the radial profile of the intensity contrast between the bar and interbar regions. Following \cite{Aguerri2000}, we defined the bar intensity as $I_{\rm bar} = I_0 + I_2 + I_4 + I_6$ and the interbar intensity as $I_{\rm ibar} = I_0 - I_2 + I_4 - I_6$. The bar region corresponds to the radial range where the bar-interbar intensity ratio is $I_{\rm bar}/I_{\rm ibar} > 0.5 [\max(I_{\rm bar}/I_{\rm ibar})-\min(I_{\rm bar}/I_{\rm ibar})]+\min(I_{\rm bar}/I_{\rm ibar})$. We adopted the FWHM of the radial profile of $I_{\rm bar}/I_{\rm ibar}$ as the bar radius. 

Then, we measured $R_{\rm bar}$ from the $\epsilon$ radial profile of the isophotal ellipses, since it traces the shape and size of the stellar orbits supporting the bar. The galaxy isophotes usually appear almost circular near the centre, while their $\epsilon$ increases up to a local maximum in the bar region and decreases outwards to a local minimum in the disc region. Following \cite{Wozniak1991}, we adopted the position of the maximum $\epsilon$ as second estimate of $R_{\rm bar}$.

Finally, we obtained a third estimate of $R_{\rm bar}$ from the analysis of the radial profile of the PA with the isophotal ellipses. The galaxy isophotes show a constant PA in the bar and disc regions. Usually, the two values are different being related to the orientation of the bar and line-of-nodes, respectively. As in Paper I, we adopted as $R_{\rm bar}$ the position where the PA changes by $\Delta {\rm PA}=5\degr$ from the PA of the ellipse with the maximum $\epsilon$ value.

We adopted for each galaxy the mean value from the three measurements and the largest deviation from the highest and lowest estimates from the mean as $R_{\rm bar}$ and corresponding upper and lower errors, respectively. This corresponds to a typical error around $30\%$. The resulting values of $R_{\rm bar}$ and error are reported in Table~\ref{tab:sample}.

\subsection{Bar pattern speed}
\label{sec:omega}

\begin{figure*}
    \centering
    \includegraphics[scale=0.65]{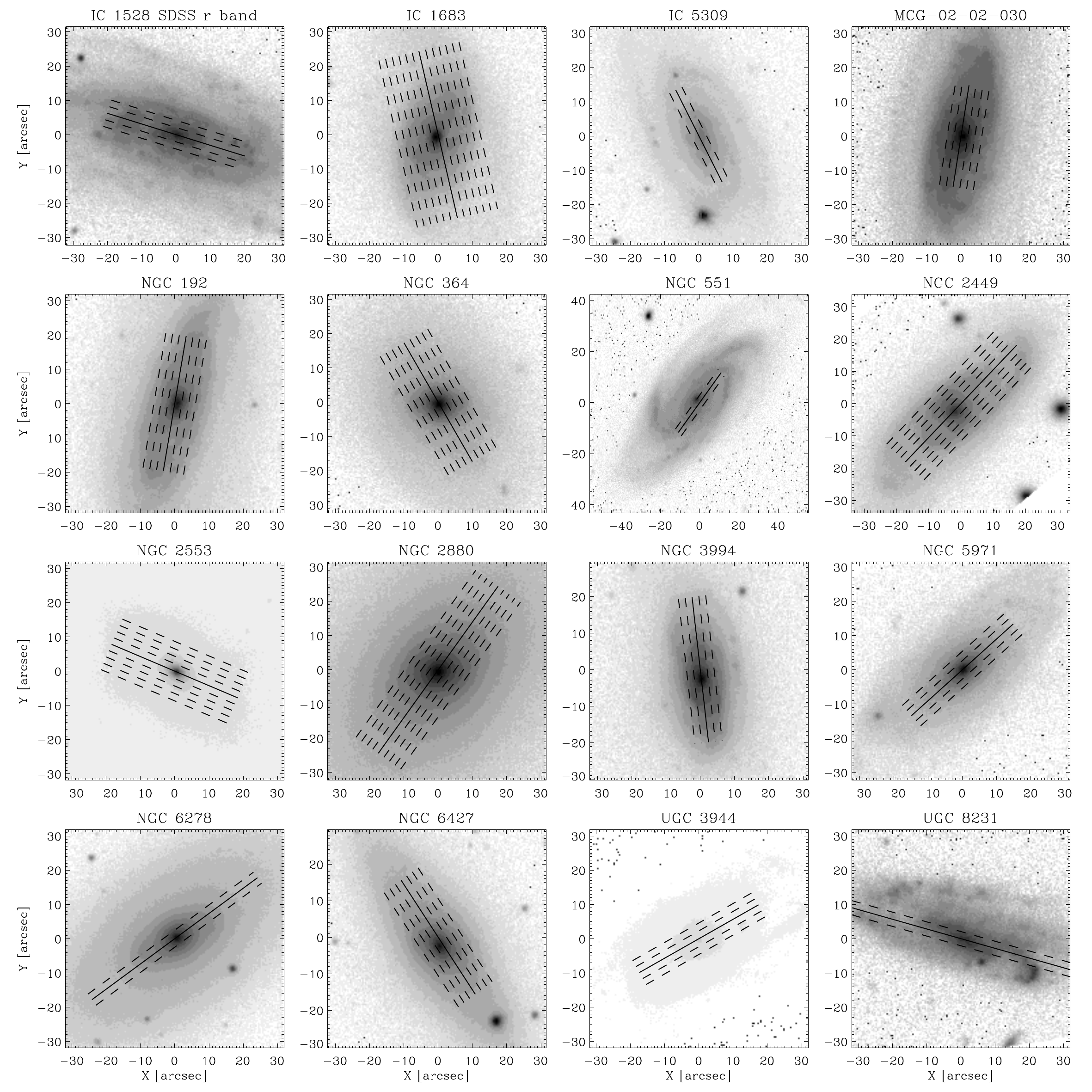}
    \caption{SDSS $r$-band images of the sample of 16 SAB galaxies successfully analysed with the TW method. The FOV is oriented with North up and East left. For each galaxy, the position and length of the pseudoslits which are parallel to the disc major axis and cross the galaxy centre (solid line) or are offset with respect to it (dashed lines) are shown.}
    \label{fig:tw_images}
\end{figure*}

We applied the TW method as outlined in Eq.~\ref{eq:tw}. For each galaxy, the pseudoslits were defined {\em a posteriori} from the CALIFA reconstructed image. We traced from three up to thirteen 1-arcsec wide pseudoslits crossing the bar and oriented with the disc PA. A minimum separation of 1 arcsec between adjacent pseudoslits was fixed to deal with independent data and minimise the impact of spatial correlations on the TW integrals. We adopted a half length of the pseudoslits in the range of $(1-4) h$, where $h$ is the exponential scalelength of the disc from the photometric decompositions of \cite{MendezAbreu2017}. In all cases we checked that the pseudoslits extended out to the axisymmetric region of the disc defined from the photometric decomposition (Fig.~\ref{fig:tw_images}).

We measured $\langle X\rangle$ in the CALIFA reconstructed image obtained by summing the CALIFA datacube along the spectral direction in a wavelength range between 4500 and 4650 \AA\ and excluding intervals severely affected by bad pixels. The spectral range was selected because it does not cover prominent emission lines. Errors on $\langle X \rangle$ in each slit were defined as the standard deviation of the $\langle X \rangle$ measured varying the slit length within the range of the constant behaviour of each integrals. The errors typically range between 0.07 and 0.15 arcsec, similar to what was found in Paper I. The convergence of the photometric integrals was checked by measuring them as a function of the coordinates of the galaxy centre and the pseudoslit length.

We measured $\langle V \rangle$ by collapsing each pseudoslit along the spatial direction and measuring the LOS velocity of the resulting spectrum in the wavelength range between 3400 \AA\ and 4750 \AA\ with the Penalized Pixel Fitting \citep[{\sc ppxf}, ][]{cappellari2004} and Gas and Absorption Line Fitting \citep[{\sc gandalf}, ][]{Sarzi2006} {\sc idl\footnote{Interactive Data Language is distributed by Harris Geospatial Solutions.}} algorithms. Since the kinematic integrals are not affected by corrupted pixels at the ends of the spectral intervals, the entire CALIFA spectral range was adopted to recover them. This approach is equivalent to using an explicit luminosity weight because the spaxels with higher signal give higher contribution in the collapsed spectrum and consequently in the $\langle V \rangle$ determination of each pseudoslit. We convolved a linear combination of $\sim 330$ stellar spectra available in the Indo-US library \citep{Valdes2004} with a line-of-sight velocity distribution (LOSVD) modelled as a truncated Gauss-Hermite series \citep{gerhard1993, vandermarel1993} via a $\chi^2$ minimization. The stellar spectra were selected to fully cover the parameter space of the effective temperature $T_{\rm eff}$, surface gravity $g$ and metallicity [Fe/H], broadened to match the CALIFA instrumental resolution. After rebinning the stellar spectra to a logarithmic scale along the dispersion direction, we redshifted them to rest frame and cropped their wavelength range to match the redshifted frame of the galaxy spectra. Moreover, a low-order multiplicative Legendre polynomial was added to correct for the different shape of the continuum of the spectra of the galaxy and optimal template due to reddening and large-scale residuals of flat-fielding and sky subtraction. 
The statistical errors on the stellar kinematic parameters were assumed to be the formal errors of the {\sc ppxf} best fit after rescaling the minimum $\chi^2$ to achieve $\chi^2_{\rm min} = N_{\rm dof} = N_{\rm d}-N_{\rm fp}$, with $N_{\rm dof}$, $N_{\rm d}$, and $N_{\rm fp}$ the number of the degrees of freedom, data points, and fitting parameters, respectively \citep{Press1992}. Errors on $\langle V \rangle$ range between 1 and 15 km s$^{-1}$. The convergence of the kinematic integrals was checked by measuring them as a function of the coordinates of the galaxy centre and pseudoslit length.

\begin{figure*}
    \centering
    \includegraphics[scale=0.65]{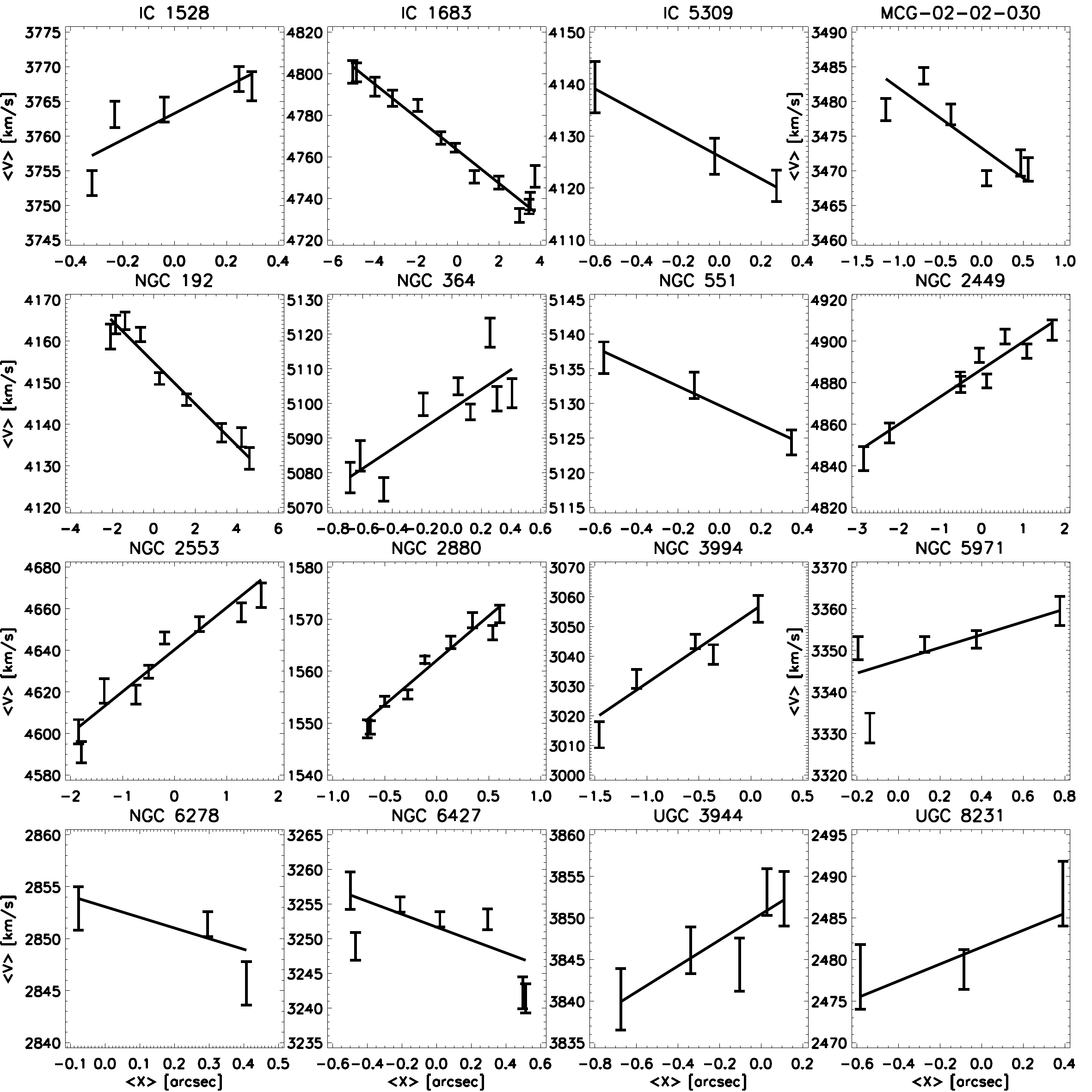}
    \caption{Bar pattern speed of the 16 SAB galaxies shown in Fig.~\ref{fig:tw_images}. For each galaxy, the kinematic integrals $\langle V\rangle$ are plotted as a function of the photometric integrals $\langle X\rangle$ and the best-fitting straight line, which has a slope equal to $\Omega_{\rm bar}\sin i$, is over-plotted.}
    \label{fig:tw_omega}
\end{figure*}

The value of $\Omega_{\rm bar} \sin i$ was obtained by fitting a straight line to the values of $\langle X\rangle$  and $\langle V \rangle$ using the {\sc idl} task {\sc fitexy} (Fig.~\ref{fig:tw_omega}). 

We calculated the value of $\Omega_{\rm bar}$ for both the PAs obtained with the photometric decomposition and isophotal analysis of the galaxy surface brightness distribution and adopting the corresponding values for $i$. For each PA, the TW integrals were measured as outlined above. For each galaxy, we obtained two estimates of $\Omega_{\rm bar}$, which are in most of the cases compatible within 1$\sigma$ of each other. Nevertheless, we have 3 cases where the $\Omega_{\rm bar}$ measurements are not consistent within the errors. The difference between the two PAs used to apply the TW method is between 2\degr and 7\degr and it explains the different results for $\Omega_{\rm bar}$. Analysing the PA radial profiles in these cases we observed the photometric decomposition was not able to effectively describe the PA of the disc. As reference result, we adopted the value of $\Omega_{\rm bar}$ obtained using the PA and $i$ defined from the radial profiles, as we already did in previous works \citep[Paper I,][]{Cuomo2019}. At the end of this analysis, 7 more galaxies were discarded (Table~\ref{tab:sample}) because of large errors on $\langle V \rangle$ ($\Delta\langle V \rangle/\langle V \rangle> 0.5$ translates into $\Delta\Omega_{\rm bar}/\Omega_{\rm bar}\sim1$), or because the presence of residual spectral features in the CALIFA datacube prevented the convergence of $\langle X \rangle$ and/or $\langle V \rangle$. 

The SDSS $r$-band images of the remaining 16 galaxies, which represent our sample of SAB galaxies successfully analysed with the TW method, are shown in Fig.~\ref{fig:tw_images}. This means that $45\%$ of the galaxies of the initial sample do not actually host a genuine bar component. The adopted values of PA and measurements of $\Omega_{\rm bar}$ are reported in Table~\ref{tab:results}. Typical errors on $\Omega_{\rm bar}$ are around $25\%$. The TW integrals and best-fitting straight lines are plotted in Fig.~\ref{fig:tw_omega}.

\begin{table*}[t!]
\centering
\caption{\label{tab:results} Properties of the bar and disc of the SAB galaxies successfully analysed with the TW method.}
\begin{tabular}{cccccc}
\hline\hline
Galaxy &  PA$_{\rm TW}$ & $\Omega_{\rm bar}$ & $V_{\rm circ}$ & $R_{\rm cr}$ & ${\cal R}$ \\
 &  [$^\circ$]  & [km s$^{-1}$ kpc$^{-1}$] & [km s$^{-1}$] & [kpc] & \\
(1) & (2) & (3) & (4) & (5) & (6) \\
\hline
IC 1528& 72.7 & $87\pm20$ & $142\pm14$ & $1.63\pm0.51$ & 0.76$^{+0.14}_{-0.22}$\\
IC 1683& 13.0 & $30.3\pm5.1$ & $191\pm45$ & $6.3\pm2.7$ & 0.71$^{+0.21}_{-0.21}$\\
IC 5309& 26.7 & $91\pm26$ & $114\pm25$ & $1.25\pm1.01$ & 0.63$^{+0.36}_{-0.45}$\\
MCG-02-02-030 & 171.1 & $43.4\pm6.5$ & $210\pm55$ & $4.83\pm2.16$ & 1.32$^{+0.36}_{-0.53}$\\
NGC 192& 170.4 & $20.9\pm2.1$ & $248.3\pm6.6$ & $11.9\pm1.9$ & 1.08$^{+0.10}_{-0.13}$\\
NGC 364& 29.9 & $120\pm31$ & $317\pm30$ & $2.63\pm1.13$ & 0.83$^{+0.22}_{-0.26}$\\
NGC 551 & 137.0 & $45\pm11$ & $202\pm43$ & $4.52\pm2.39$ & 1.17$^{+0.39}_{-0.71}$\\ 
NGC 2449& 136.4 & $40.7\pm5.5$ & $236.9\pm2.6$ & $5.84\pm0.99$ & 1.27$^{+0.11}_{-0.14}$\\
NGC 2553& 67.0 & $68.1\pm9.8$ & $269\pm34$ & $3.95\pm0.91$ & 0.515$^{+0.077}_{-0.110}$\\
NGC 2880& 144.6 & $190\pm28$ & $209\pm15$ & $1.09\pm0.36$ & 0.74$^{+0.15}_{0.19}$\\
NGC 3994& 6.9 & $119\pm27$ & $226.4\pm5.5$ & $1.90\pm0.67$ & 1.06$^{+0.22}_{-0.31}$\\
NGC 5971& 132.0 & $55\pm15$ & $226\pm16$ & $4.07\pm1.96$ & 0.56$^{+0.15}_{-0.32}$\\
NGC 6278& 126.4 & $92\pm28$ & $279\pm13$ & $3.05\pm1.06$ & 1.07$^{+0.26}_{-0.25}$\\
NGC 6427& 34.7 & $46\pm10$ & $245\pm21$ & $5.3\pm3.6$ & 2.8$^{+1.0}_{-1.8}$\\
UGC 3944& 119.6 & $62\pm22$ & $148\pm30$ & $2.39\pm11.6$ & 1.28$^{+3.8}_{-5.7}$\\
UGC 8231& 74.2 & $58\pm31$ & $136\pm27$ & $2.3\pm5.3$ & 1.01$^{+1.6}_{-2.0}$\\
\hline
\end{tabular}
\tablefoot{(1) Galaxy name. (2) Adopted value of the position angle of the pseudoslits for the TW analysis. (3) Bar pattern speed. (4) Disc circular velocity. (5) Bar corotation radius. (6) Bar rotation rate.}
\end{table*}

\subsection{Corotation radius and bar rotation rate}
\label{sec:rate}

Although the TW method does not need any modelling to derive $\Omega_{\rm bar}$, we need some assumptions to derive the circular velocity $V_{\rm circ}$ and consequently $R_{\rm cr}$ and ${\cal R}$. 

We obtained $V_{\rm circ}$ from the maps of stellar LOS velocity and velocity dispersion in the disc region provided by \cite{FalconBarroso2017} using the asymmetric drift equation \citep{Binney1987} and following the prescriptions of \cite{Debattista2002} and \cite{Aguerri2003}. In particular, we assumed an exponential radial profile with the same scalelength for the radial, azimuthal, and vertical components of the velocity dispersion $\sigma_R, \sigma_\phi, \sigma_z$, the epicyclic approximation and a constant $V_{\rm circ}$ resulting in $\sigma_\phi/\sigma_R=1/\sqrt{2}$, and a value for $\sigma_z/\sigma_R$ depending on the morphological type \citep[][Paper I]{Gerssen2012}.

\begin{figure}
    \centering
    \includegraphics{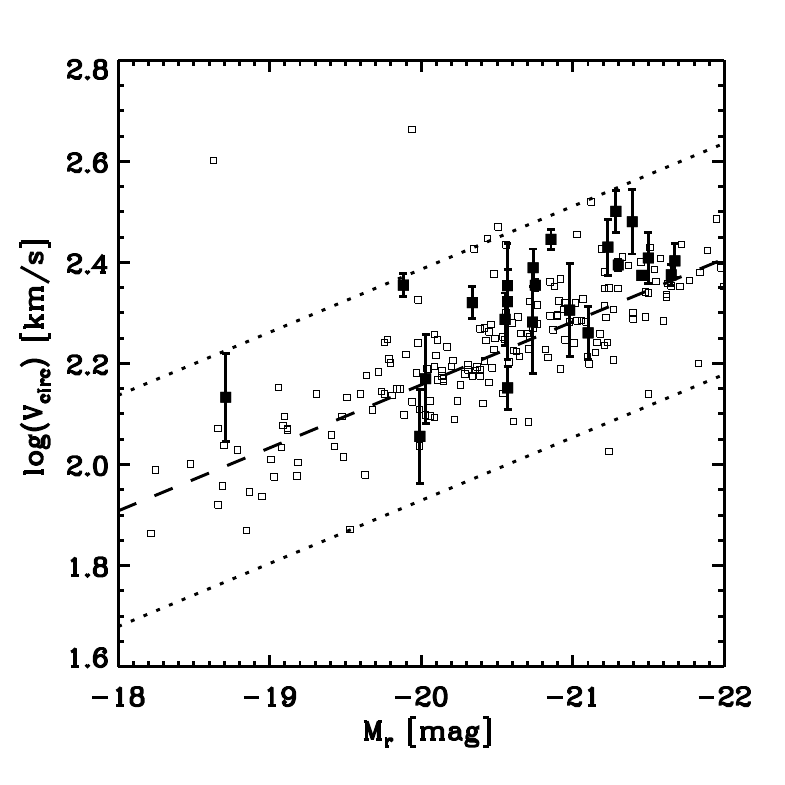}
    \caption{Circular velocity as a function of absolute $r$-band magnitude of
    the SAB galaxies successfully analysed with the TW method (filled squares) and the galaxy sample 
    of \cite{Reyes2011} (open squares). The dashed line is the best-fitting relation of \cite{Reyes2011} and the dotted lines bracket the region of $3\sigma$ deviation in $\log{(V_{\rm circ})}$.}
    \label{fig:tully_fisher}
\end{figure}

We checked the reliability of our circular velocities by performing a comparison with the $V_{\rm circ}$ values predicted by the Tully-Fisher relation of \cite{Reyes2011} and with those obtained by \cite{Leung2018} with different dynamical models than ours. Our galaxies are consistent within the $3\sigma$ scatter of the relation between the circular velocity and absolute SDSS $r$-band magnitude calculated by \cite{Reyes2011} for a sample of $\sim200$ nearby SDSS galaxies (Fig.~\ref{fig:tully_fisher}). Eight galaxies in our sample are common with the subsample of $\sim50$ CALIFA galaxies studied by \cite{Leung2018}. Our values of $V_{\rm circ}$ are in agreement within the errors with their values.

We calculated $R_{\rm cr} = V_{\rm circ}/\Omega_{\rm bar}$ from the asymmetric-drift circular velocity and TW bar pattern speed and we derived the ratio between the corotation and bar radius ${\cal R} = R_{\rm cr}/R_{\rm bar}$. The values of $V_{\rm circ}$, $R_{\rm cr}$ and ${\cal R}$ are reported in Table~\ref{tab:results}.

\section{Results}
\label{sec:results}

\subsection{Ultrafast bars}

Two galaxies with a TW-measured $\Omega_{\rm bar}$ ($13\%$) host an ultrafast bar  having ${\cal R}<1$ at $95\%$ confidence level (Table~\ref{tab:results}). This ${\cal R}$ regime corresponds to bars extending beyond $R_{\rm cr}$, which are expected to rapidly dissolve. 

In Paper I, we explored possible explanations for measuring ${\cal R}<1$ with the TW method, which are including obtaining the wrong estimate of $R_{\rm bar}$ and/or $R_{\rm cr}$, the application to objects which do not meet all the TW requirements, or the presence of multiple pattern speeds associated with the main bar, the spiral arms, and a nuclear bar. In order to address these issues, we obtained $R_{\rm bar}$ with three different and independent methods. In some cases, these estimates are quite different from each other, but this reflects on the adopted error on $R_{\rm bar}$ which we defined as the largest difference between the mean value and the three measurements. On the other hand, $R_{\rm cr}$ depends on both $V_{\rm circ}$ and $\Omega_{\rm bar}$. The circular velocity was obtained using the asymmetric drift correction and the resulting values are consistent with the predictions of the Tully-Fisher relation \citep{Reyes2011} and previous measurements based on different dynamical models \citep{Leung2018}. In \citet{Cuomo2019} we have shown that ${\cal R}<1$ could be the result of a wrong estimate of the disc PA when the PA radial profile does not present a constant trend in the disc region. This is the reason why we checked the constancy of the profiles and finally used the PA from the photometric analysis to recover $\Omega_{\rm bar}$. A slope change with radius of the straight-line fitting $\langle X \rangle$ and $\langle V \rangle$ is interpreted as due to components rotating with a different pattern speed with respect to the main bar \citep{Corsini2003, Maciejewski2006, Meidt2009}. This change is observed in IC~1683, NGC~2553, and NGC~6427 (Fig.~\ref{fig:tw_omega}), although not all of them host an ultrafast bar (Table~\ref{tab:results}).

We run a Kolmogorov-Smirnov (KS) test with the {\sc idl} procedure {\sc kstwo} to verify if there are statistical differences between the distributions of morphological type, redshift, and absolute $r$-band magnitude of the initial sample of 29 {\em bona fide} SAB galaxies, the sample of 16 SAB galaxies successfully analysed with the TW method, and the sample of 14 SAB galaxies without an ultrafast bar (Fig.~\ref{fig:histograms}). Since we found no significant difference at a very high confidence level ($>95\%$) between the properties of the three samples, we decided to not consider further the ultrafast bars.

\subsection{Bar properties in weakly and strongly barred galaxies}
\label{sec:weak_strong}

\begin{figure*}
    \centering
    \includegraphics[scale=0.8]{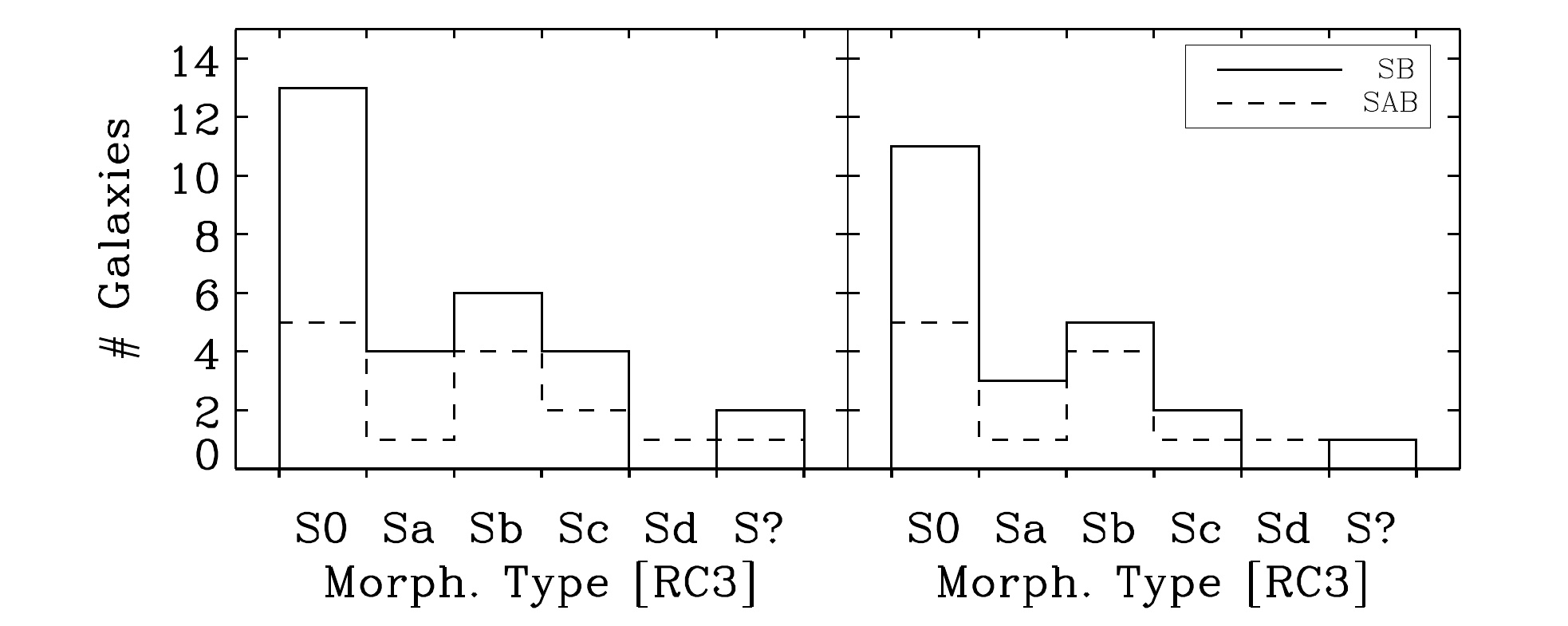}
    \caption{Distributions of the morphological type of the SB (solid line) and SAB (dashed line) galaxies including ({\em left panel}) and excluding ({\em right panel}) the ultrafast bars.}
    \label{fig:morphology}
\end{figure*}

Our goal is to compare the bar properties of a sample of SB and SAB galaxies with a TW-measured $\Omega_{\rm bar}$ as well as of their host galaxies. To perform an effectively comparison, the different bar properties have to be derived using similar methodology within the sample. To this aim, we added ESO-139-G0009 \citep{Aguerri2003} to our sample of SAB galaxies without an ultrafast bar and as a comparison sample of SB galaxies, we collected 17 (\citealt{Debattista2004}, Paper I, \citealt{Cuomo2019}) and 15 galaxies \citep{Merrifield1995, Gerssen1999, Debattista2002, Aguerri2003, Corsini2003, Gerssen2003, Corsini2007, Treuthardt2007} with $\Omega_{\rm bar}$ measured with the TW method from the stellar kinematics obtained with integral-field and long-slit spectroscopy, respectively. Only three out of 32 SB galaxies host an ultrafast bar ($9\%$). In Fig.~\ref{fig:morphology} we show the distributions of morphological type of the SAB and SB galaxies including or excluding the ultrafast bars. The remarkably large number of SB0 galaxies is an effect of the selection bias due to the application of the TW method to early-type disc galaxies with a low dust and gas content \citep[see][for a review]{Corsini2011}.

\begin{figure*}
    \centering
    \includegraphics{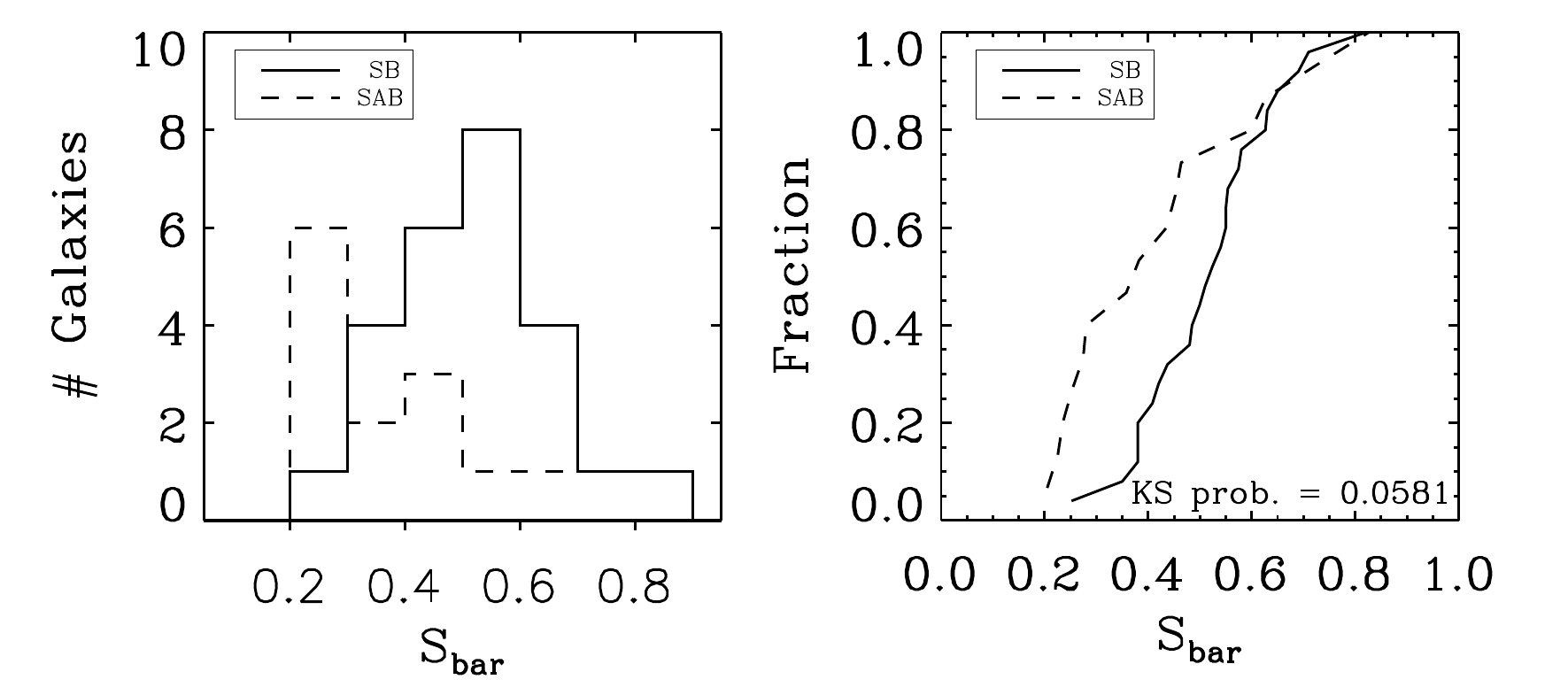}
    \caption{Distribution of the bar strength of the SB (solid line) and SAB (dashed line) galaxies excluding the ultrafast bars ({\em left panel}) and cumulative distributions of SB (solid line) and SAB (dashed line) galaxies without any ultrafast bars as a function of bar strength ({\em right panel}).}
    \label{fig:visually_sb_sab}
\end{figure*}

We investigated the distributions of absolute $r$-band magnitude and bar properties of SAB and SB galaxies without an ultrafast bar, respectively. For each parameter, we performed a KS test to look for statistically significant differences between the two samples. We found that the bars in SAB galaxies are similar to those of SB for all the explored parameters, in particular the two samples have similar bar strengths. This is due to the fact that visually-classified SAB galaxies are contaminated by strong bars while comparing the strength of the bar. In fact, the mean strength value of the SAB galaxies is $\langle S_{\rm bar} \rangle= 0.42\pm0.18$. On the other hand, visually-classified SB galaxies may host weak bars in term of the strength (Fig.~\ref{fig:visually_sb_sab}). This point towards the fact that a visual separation between weak and strong bars does not correspond to classify the galaxies according to the bar strength.

Since a quantitative distinction between strong and weak bars is not defined \citep{Athanassoula2003, Athanassoula2013, Vera2016, Kruk2018}, we decided to split our full sample of 46 galaxies using a quantitative criterion based on the strength, if available. The chosen limiting value is $S_{\rm bar}=0.4$, which corresponds to include $50\%$ of visually-classified SAB galaxies in the new quantitative defined SAB sample and to have enough objects in SB and SAB samples to perform some significant statistics. Speaking about strong and weak bars, from now on we refer to quantitative SB and SAB galaxies, because their definition is based on the strength.

\begin{figure*}
    \centering
    \includegraphics[scale=0.6]{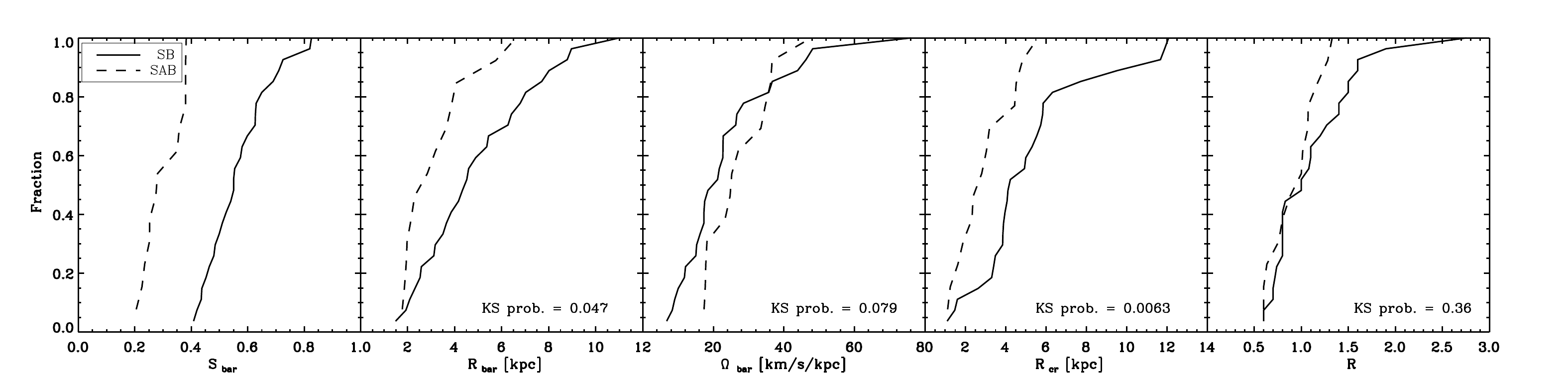}
    \caption{Cumulative distributions of SB (solid line) and SAB (dashed line) galaxies without an ultrafast bar as a function of bar radius, bar strength, bar pattern speed, corotation radius, and bar rotation rate ({\em from left to right panel}). The significance level of the KS test is given in each panel.}
    \label{fig:bar}
\end{figure*}

\begin{figure}
    \centering
    \includegraphics{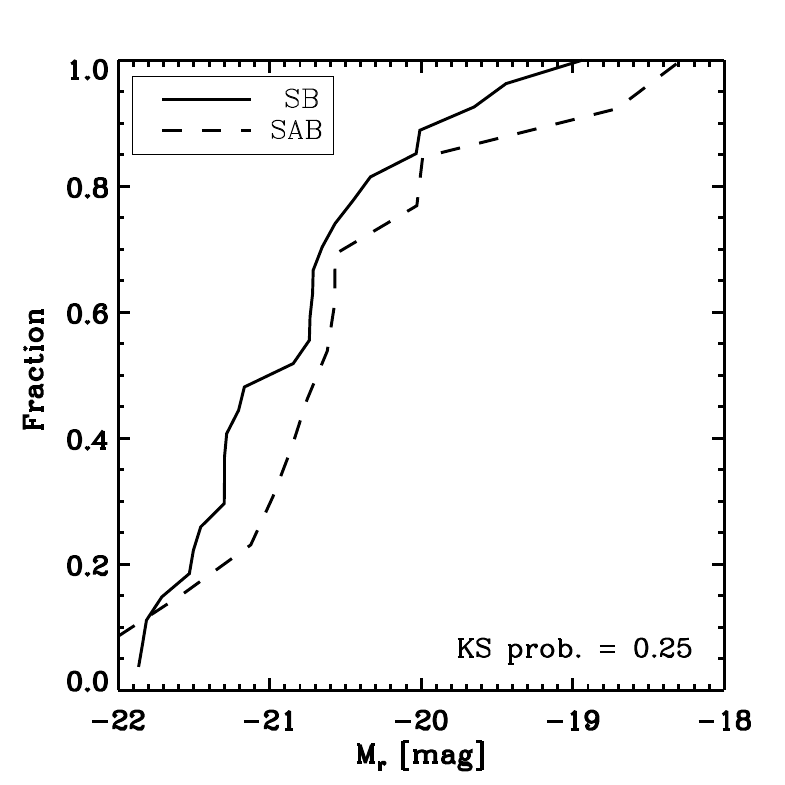}
    \caption{Cumulative distributions of SB (solid line) and SAB (dashed line) galaxies without an ultrafast bar as a function of absolute $r$-band magnitude. The significance level of the KS test is given.}
    \label{fig:magnitude}
\end{figure}

We investigated the distributions of bar properties and absolute $r$-band magnitude of the quantitative defined 13 SAB and 27 SB galaxies classified through the bar strength and without an ultrafast bar (Figs.~\ref{fig:bar} and \ref{fig:magnitude}). For each parameter, we performed a KS test to look for statistically significant differences between the two samples. We confirmed that the bars of SAB galaxies are weaker that those of SB galaxies although their hosts have the same luminosity distribution. In addition, we found at a very high confidence level ($>99\%$) that weak bars are shorter and have smaller $R_{\rm cr}$ with respect to their strong counterparts. On the other hand, SAB and SB galaxies display similar distributions of $\Omega_{\rm bar}$ and the bar rotation rate. We repeated the analysis using the results obtained with the PA from the photometric decomposition and we obtained similar results. The relations between bar and galaxy properties are investigated in the following section.

\subsection{Bulge and disc properties in weakly and strongly barred galaxies}
 
We analysed the relations between the bar parameters and the bulge and disc properties of the SAB and SB galaxies. We recovered the bulge-to-total luminosity ratio $B/T$, bulge S\'ersic parameter $n$ and effective radius $R_{\rm e}$ of the bulge, and scalelength $h$ of the disc from \cite{MendezAbreu2017} for the CALIFA galaxies and from the quoted papers for the other galaxies. 

We performed a KS test on the bulge properties and found that SB and SAB galaxies present at a high significance level the same distributions for $n$ and $R_{\rm e}$ of the bulges, but different distributions for $B/T$. Moreover, two SAB galaxies turned out to be bulgeless, while the result of KS test on bulge properties remain the same even when discarding these two objects. This analysis suggests that bulges of SAB and SB galaxies present similar properties but a different contribution to the total light of their host galaxies.

\begin{figure*}
    \centering
    \includegraphics[scale=0.75]{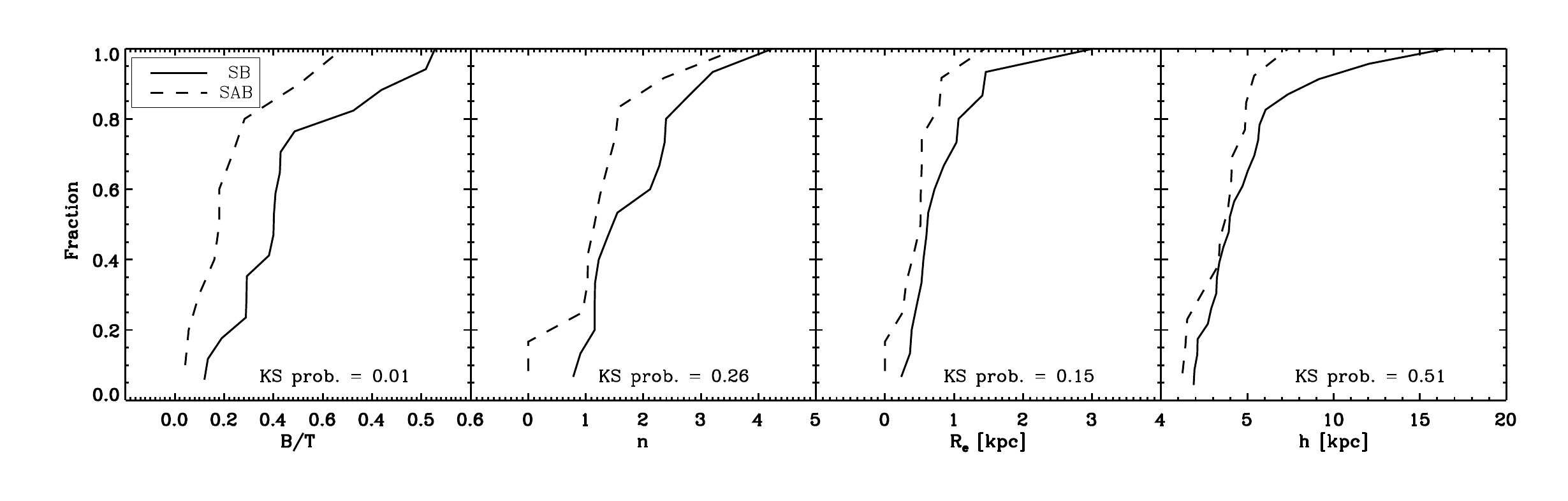}
    \caption{Cumulative distributions of SB (solid line) and no bulgeless SAB (dashed line) galaxies without an ultrafast bar as a function of bulge-to-total luminosity, bulge S\'ersic index, and bulge effective radius; cumulative distributions of SB and SAB galaxies without an ultrafast bar as a function of disc scalelength ({\em from left to right panel}). The significance level of the KS test is given in each panel.}
    \label{fig:bulge}
\end{figure*}

\begin{figure}
    \centering
    \includegraphics[scale=1.2]{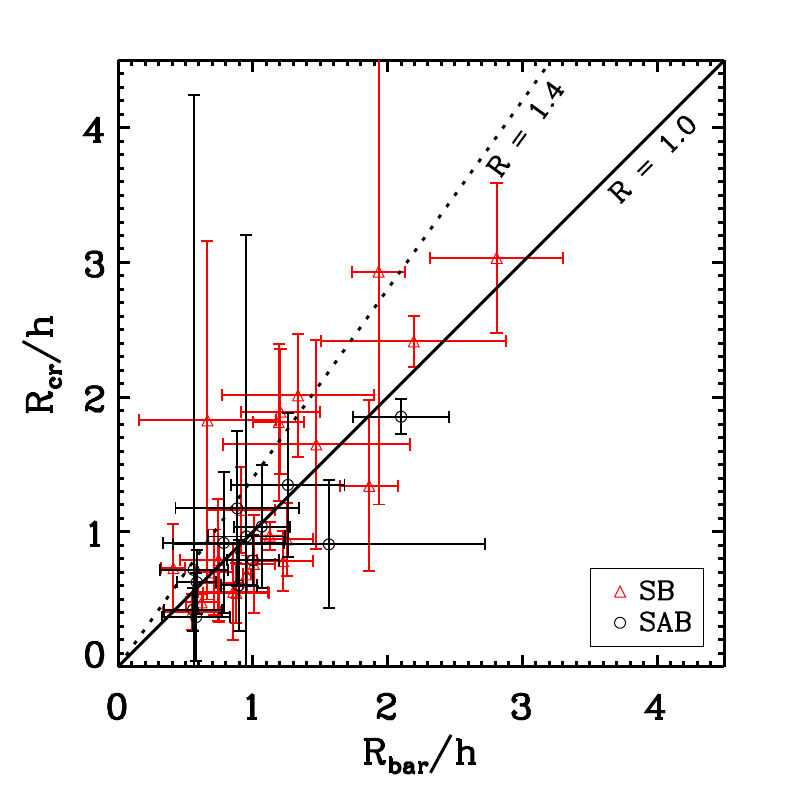}
    \caption{Ratio between the corotation radius and disc scalelength as a function of the ratio between the bar radius and disc scalelength for SB (red triangles) and SAB (black circles) galaxies without an ultrafast bar. The solid and dashed lines mark ${\cal R}=1.0$ and 1.4, respectively.}
    \label{fig:disc}
\end{figure}

The discs of SAB and SB galaxies are also similar to each other (Fig.~\ref{fig:bulge}). To investigate the disc regions hosting weak and strong bars, we measured the ratios between $R_{\rm cr}$ and $h$ and between $R_{\rm bar}$ and $h$ for the SAB and SB galaxies (Fig.~\ref{fig:disc}). Most of the bars and corotation radii of both galaxy samples are confined within or are close to their disc scalelength since the $R_{\rm cr}/h$ and $R_{\rm bar}/h$ typically range between 1 and 1.5. In particular, for SAB galaxies both $R_{\rm bar}/h$ and $R_{\rm cr}/h$ are smaller than 1.5 (except for one outlier, whose $S_{\rm bar}$ lies near the limiting value adopted to split the sample), while $30\%$ of SB galaxies are characterised by $R_{\rm cr}/h$ and $R_{\rm bar}/h$ larger than 1.5. The ratio of $R_{\rm cr}/h$ and $R_{\rm bar}/h$ corresponds to ${\cal R}$, which ranges between 1 and 1.4 corresponding to the fast bar regime.

\section{Discussion and conclusions}
\label{sec:conclusions}

In this work, we obtained the bar properties of a sample of 29 {\em bona fide} SAB galaxies by analysing the $r$-band images available from the SDSS survey and stellar kinematic maps obtained from the CALIFA survey (Table~\ref{sec:sample}). The galaxies were selected to have an intermediate inclination, a bar elongated in between the minor and major axes of the disc and to be morphologically and kinematically undisturbed. The sample galaxies have morphological types ranging from S0 to Scd, with redshifts between 0.005 and 0.30 and absolute $r$-band total magnitudes from $-18.5$ to $-23.0$ mag.

We derived the bar radius $R_{\rm bar}$ in the deprojected images of the galaxies by measuring the bar-interbar intensity ratio obtained from the Fourier analysis of the surface brightness distribution, the location of the maximum in the $\epsilon$ radial profile and the behaviour of the PA radial profile of the ellipses fitting the galaxy isophotes. At the same time, we measured the bar strength $S_{\rm bar}$ from the Fourier analysis. Despite the stringent criteria we adopted for the selection, we discarded six galaxies because they turned out not to host a clear bar component. The $m=2$ Fourier component did not show the amplitude peak with a constant phase angle typical of barred galaxies, while the large odd components revealed the presence of non-axisymmetric structures other than a bar.

We applied the TW method to obtain the bar pattern speed $\Omega_{\rm bar}$ from the CALIFA datacubes. This study represents the third effort to apply the TW method to a large sample of galaxies based on integral-field spectroscopy, and the first one including SAB galaxies. To this aim, we measured the luminosity-weighted mean position and LOS velocity of the stars across the bar in several pseudoslits parallel to the disc major axis. We rejected 7 more galaxies because of the poor correlation, or the large errors, or the non-convergence of the TW integrals. This means that 13 galaxies of the sample ($45\%$), which were morphologically classified as weakly barred from a visual inspection, do not actually host a genuine bar component or the central elongated structure is not in rigid rotation. For the remaining 16 SAB galaxies, we derived the corotation radius $R_{\rm cr}$ from the circular velocity obtained by applying the asymmetric drift correction to the stellar kinematics and the bar rotation rate ${\cal R}$ as the ratio between $R_{\rm cr}$ and $R_{\rm bar}$. All the measured SAB bars are consistent with being fast within the errors ($1<{\cal R}<1.4$), except for two of them which are ultrafast (${\cal R}<1$) at $95\%$ confidence level (Table~\ref{tab:results}) and were not considered further. Although several ultrafast bars have been found with the TW method using integral-field spectroscopic data (Paper I, \citealt{Guo2019}), their dynamics is not yet fully explained and requires a deeper analysis both from an observational and theoretical point of view. 

We built a comparison sample of SB galaxies with TW-based $\Omega_{\rm bar}$ from the literature (Fig.~\ref{fig:morphology}). We split the entire sample of 46 barred galaxies (visually-classified SB + SAB) analysed with the TW method so far according to the strength (if available) of the bar and excluding the ultrafast galaxies. The value $S_{\rm bar}=0.4$ is adopted to provide a quantitative definition of SAB and SB galaxies, and the final sample includes 13 quantitative SAB and 27 quantitative SB galaxies. The SAB galaxies host weaker and shorter bars with smaller corotations than bars of SB galaxies. In the end, both SAB and SB galaxies have similar large pattern speeds and bar rotation rates and therefore host {\em fast} bars (Fig.~\ref{fig:bar}). After checking that the two samples do have similar absolute total magnitudes, we excluded that this result is due to a bias in the distribution of their luminosities (Fig.~\ref{fig:magnitude}). Since SAB galaxies, similarly to SB galaxies, host {\em fast} bars, we can exclude that their formation was tidally triggered by a past interaction with a companion. The numerical simulations by \cite{MartinezValpuesta2017} and \cite{Lokas2018} show that tidally-induced bars suffer a steady weakening across their evolution but their rotation rate is always in the slow regime. Our SAB sample includes many early-type disc galaxies (Fig.~\ref{fig:morphology}), which were found to host fast bars in earlier studies \citep[e.g.,][]{Rautiainen2008, Font2017}. However, we did not find a significant correlation between ${\cal R}$ and morphological type because of the small number statistics.

Since one of the most promising and often advocated causes of bar weakening is the presence of a central mass concentration, we investigated the relation between the presence of a weak/strong bars and the bulge properties of the host galaxy. We did not find any significant difference in the Se\'rsic index $n$ and effective radius $R_{\rm e}$ of the bulges of SAB and SB galaxies. Instead we find a lower $B/T$ ratio in SAB galaxies. Moreover, we found two bulgeless SAB galaxies. A similar result was found by \cite{Abraham2000}, who showed that SAB galaxies are less concentrated that their SB counterparts. Therefore, we conclude that the presence of a prominent bulge does not necessarily imply the bar weakening. Moreover, we clearly found that $\Omega_{\rm bar}$ of weak and strong bars is similar, as previously suggested by measurements with other methods \citep{Font2017}. This allowed us to discard the dissolution scenarios, which always predict an increase in $\Omega_{\rm bar}$ while the bar is losing strength and dissolving, regardless of different causes of dissolution, such as the presence of central mass concentration, shape of DM halo, or gas accretion \citep{Athanassoula2003,Bournaud2005,Athanassoula2005}. \cite{Laurikainen2013} suggested that bulges in the early-type SB galaxies are built by bars, while those in the SAB galaxies are possibly the end result a several accretion events that occurred before the bar formation, prescribing different values for the $n$ index. In our sample this formation mechanism is not supported because we observe the same distribution of $n$ in SB and SAB galaxies. We can not further investigate the bulge type in SB and SAB galaxies, because the S\'ersic index, $n$, does not provide a clear separation between classical and pseudobulges  and a variety of spectroscopic and photometric diagnostics including the bulge intrinsic shape is needed \citep{Costantin2017, Costantin2018,MendezAbreu2018bis}.

We explored the relation between the presence of a weak/strong bar and the disc scalelength of the host galaxy. We found that weak bars are all hosted in the inner parts of discs, because most of SAB galaxies have both $R_{\rm bar}/h$ and $R_{\rm cr}/h$ smaller than 1.0 and in all SAB galaxies these ratios are smaller than 1.5, except for one outlier. We observed a larger spread of $R_{\rm bar}/h$ and $R_{\rm cr}/h$ for SB galaxies, with a clear tail to values larger than 1.5 (Fig.~\ref{fig:disc}). 

A generalized picture for bar formation and evolution may be summarized as follows. A bar in the early stage of evolution extends out to the corotation (${\cal R}\sim1$) and presents a high value of $\Omega_{\rm bar}$. Then, both $R_{\rm bar}$ and $R_{\rm cr}$ increase as a consequence of the angular momentum exchange between the bar and other galactic components, while $\Omega_{\rm bar}$ decreases. At some point during the evolution, the corotation reaches the disc region where the star density is too low to further feed the bar. From this moment, $R_{\rm cr}$ increases more than $R_{\rm bar}$ and the rotation rate is expected to enter the slow regime (${\cal R}>1.4$) \citep{Debattista2006,Athanassoula2013}.

In this scenario, SAB galaxies with small $R_{\rm bar}/h$ and $R_{\rm cr}/h$ could be young bars, while SB galaxies with large $R_{\rm bar}/h$ and $R_{\rm cr}/h$ could be old bars. However, SB and SAB galaxies present similar value of $\Omega_{\rm bar}$ and none of the bars analysed in this work or in previous TW-based works are unambiguously located in the slow regime. Moreover, it is very unlikely to catch a bar in its early phase of evolution because the bar formation phase is very short. All these evidences suggest that SAB galaxies are dynamically evolved systems which did not exchange as much angular momentum as the SB galaxies and their hosting bars have not grown, while the paucity of slow bars remains unexplained. To confirm this scenario, further observations, dynamical modelling, and numerical simulations focused onto SAB galaxies are required because it is known that the exchange of angular momentum between the bar and other components depends on several parameters including the DM central concentration \citep[e.g.][]{Debattista2000}, initial gas fraction and halo triaxiality \citep[e.g.][]{Athanassoula2013}, disc thickness \citep[e.g.][]{Klypin2009}, and stellar mass distribution and/or weak interactions not always clearly visible in the velocity fields \citep[e.g.][]{Salak2019}.

\begin{acknowledgements}
We thank the anonymous referee for the constructive suggestions. We are grateful to M. Bureau, P. T. de Zeeuw, J. Falc\'on-Barroso, F. Pinna, L. Costantin, I. Pagotto and M. Rubino for their valuable comments. VC acknowledges support from the Fondazione Ing. Aldo Gini and thanks the Instituto de Astrof\'\i sica de Canarias and the Universidad de la Laguna for hospitality during the preparation of this paper. VC, EMC, and AP are sup ported by Padua University through grants DOR1715817/17, DOR1885254/18, DOR1935272/19, and BIRD164402/16. JALA and JMA are supported by the Spanish MINECO grants AYA2017-83204-P and AYA2013-43188-P. VPD is supported by STFC Consolidated grant ST/R000786/1. 
\end{acknowledgements}

\bibliographystyle{aa}

\end{document}